%% file: main.tex
\documentclass[sigconf,proceedings,]{acmart}
\usepackage{xcolor}

\newif\ifdraft
\drafttrue

\ifdraft
 \newcommand{\hm}[1]{{\textcolor{magenta}{ ***Heng: #1 }}}
\else
 \newcommand{\hm}[1]{}
\fi

\AtBeginDocument{%
  }

\setcopyright{acmlicensed}
\copyrightyear{2018}
\acmYear{2018}
\acmDOI{XXXXXXX.XXXXXXX}

\acmConference[Conference acronym 'XX]{Make sure to enter the correct
  conference title from your rights confirmation emai}{June 03--05,
  2018}{Woodstock, NY}
\acmISBN{978-1-4503-XXXX-X/18/06}

\begin{document}

\title{Connecting Large Language Model Agent to High Performance Computing Resource}

\author{
Heng Ma$^{1,2}$,
Alexander Brace$^{1,2}$, 
Carlo Siebenschuh$^{1,2}$,
Greg Pauloski$^{2}$,
Ian Foster$^{1,2}$,
Arvind Ramanathan$^{1,2*}$
}

\affiliation{
$^{1}$Data Science and Learning Division, Argonne National Laboratory, Lemont, IL, USA. \\
$^{2}$Computer Science Department, University of Chicago, Chicago, IL, USA. \\
Contact author: ramanathana@anl.gov
\country{}
}

\renewcommand{\shortauthors}{Ma et al.}

\begin{abstract}
The Large Language Model agent workflow enables the LLM to invoke tool functions to increase the performance on specific scientific domain questions. 
To tackle large scale of scientific research, it requires access to computing resource and parallel computing setup. 
In this work, we implemented Parsl to the LangChain/LangGraph tool call setup, to bridge the gap between the LLM agent to the computing resource. 
Two tool call implementations were set up and tested on both local workstation and HPC environment on Polaris/ALCF. 
The first implementation with Parsl-enabled LangChain tool node queues the tool functions concurrently to the Parsl workers for parallel execution. 
The second configuration is implemented by converting the tool functions into Parsl ensemble functions, and is more suitable for large task on super computer environment. 
The LLM agent workflow was prompted to run molecular dynamics simulations, with different protein structure and simulation conditions. 
These results showed the LLM agent tools were managed and executed concurrently by Parsl on the available computing resource. 
The source code is available at \url{https://github.com/ramanathanlab/SimAgent}. 

\end{abstract}




\maketitle

\input{introduction}
\input{method}
\input{results}
\input{conclusion}


\bibliographystyle{ACM-Reference-Format}
\bibliography{ref}










\end{document}

%% file: introduction.tex
\section{Introduction}
The Large Language Model (LLM) agent workflow features an interactive environment that enables the LLM to operate under different contexts to perform tasks autonomously or semi-autonomously.~\cite{langgraph,Swarm_openai,wu2023autogen, wang2023plan}
The workflow leverages the LLM's abilities of natural language contextual understanding and generative power. 
It takes in the user prompt and passes it through the workflow path of multiple LLM agents that collaborate with each other to complete the task. 
This kind of workflow can be implemented via different packages with distinct focuses, such as LangGraph for complex multistep AI workflows, Swarm/OpenAI for distributed AI collaboration and Autogen/Microsoft for software/code development tasks.~\cite{wu2023autogen}

For example, LangGraph presents a graph-based AI workflow, where the LLMs act as nodes and are connected with directed or conditional edges defining the data flow and relationship between different models.  
The workflow is compatible with the LangChain codebase, that integrates various APIs and models, allowing multi-modal applications. 
It also provides a high degree of control over the workflow dependencies. 
Such features make it the suitable framework for complex multistep tasks, such as scientific research and decision-making.

Compared to a single LLM, the LLM agent workflow can break down complex multistep tasks into manageable subtasks and delegates them to LLM agents with complementary capabilities. 
Each agent in the workflow can be prompted in playing a different role in the collaborative activities. 
This framework outperforms a single LLM in efficiency, extensibility, and adaptability in complex scientific research tasks.

Another important feature is the tool calling capability of the latest LLM models, such as OpenAI, Anthropic, Google Gemini and Mistral. 
The tool calling functionality extends the LLM capability beyond text generation, allowing it to have access to external tool functions, APIs and computer systems. \cite{schick2024toolformer}
It enables the model to access real-time data, perform specialized tasks, and integrate with complex workflow. 
This integration also helps to overcome the struggles of LLM natural language contextual understanding in the mathematical precision and specialized scientific domain tasks.

One of its most notable applications is to incorporate well-defined scientific methods, such as arithmetic functions, finite element analysis, etc. 
It gives the model access to these methods, which provide more reliable results than the text completion for the logic-base tasks for LLM. 
Rather than understanding the underlying science of a task, the LLM invokes a tool function that offers a principled solution. This approach aligns with the strengths of LLMs, enabling them to act as the connective tissue between diverse scientific applications, thereby supercharging computational workflows for science.
It combines the LLM language contextual understanding and scientific domain knowledge from the provided methods in tool function form, making it a more reliable and versatile AI approach for scientific research.

However, LLMs and many scientific workflows are computationally demanding and require parallel execution on high-performance computing (HPC) platforms such as Aurora and Polaris at the Argonne Leadership Computing Facility (ALCF). 
The computing resources of these platforms are managed through job schedulers such as the Simple Linux Utility for Resource Management (Slurm), Portable Batch System (PBS), or Load Sharing Facility (LSF). 
The scheduler requires a submission script on the login node to submit jobs to the computing queue. 
For an LLM agent to call a function on HPC, it will require such a submission script specifying the computing configurations. 
This work aims to integrate the parallel execution on local/HPC computing resource to the LLM tool calling functions via Parsl, enabling the LLM agent to carry out large scale scientific research.


%% file: method.tex
\section{Methods}
To build the framework that allows LLM agent to leverage the HPC computing power, it requires the integration of an LLM agent workflow package and HPC resource/workload manager.

\subsection{LangGraph: The graph-based AI workflow}
LangGraph was adopted to build the AI test cases in this work as its focus on complex multistep AI workflow.~\cite{langgraph}
It features a graph-based workflow and execution, which provides clarity and complex dependency management. 
This graph-based architecture uses the nodes to represent a task or LLM model, and edges to define relationships and dataflow. 
The integration of LangChain framework gives the user access to various LLM models and customized tool functions.~\cite{langchain}

In a LangGraph workflow, the graph state keeps track of all the previous events in the workflow, and coordinates the activation of the next agent node.
The active agent responds to the current graph state and updates it according to the output from the LLM model or tool functions. 
By carrying out such operation iteratively, the workflow optimizes and updates the LLM responses until the result is evaluated by the reasoning agent to be acceptable for the final output. 
This gives the workflow traceability and logging capability to make it auditable and transparent, which are essential for scientific research tasks.

\subsection{Workflow setup}
\label{sec:wf_setup}

\begin{figure*}
    \centering
    \includegraphics[width=.8\textwidth]{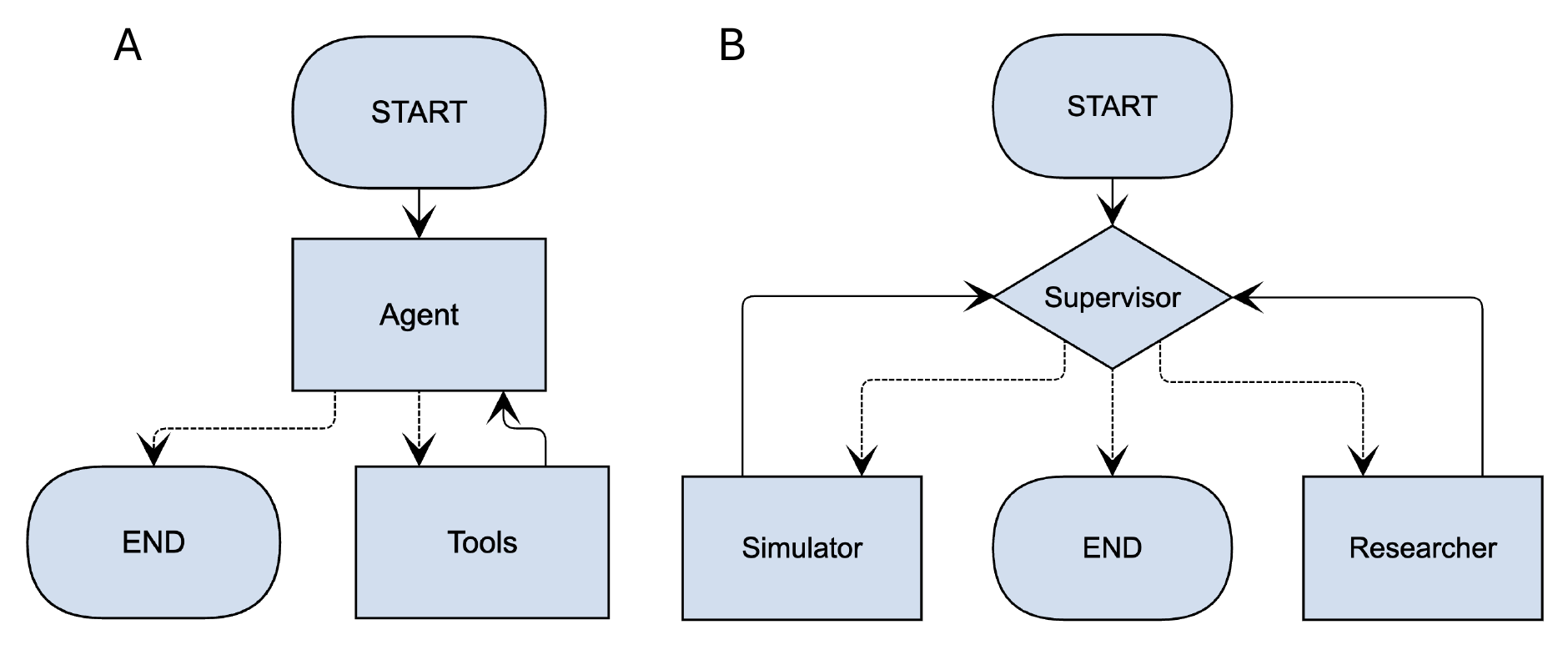}
    \caption{The LangGraph workflows, (A) workflow 1 and (B) workflow 2. The START node takes in the prompt from the user, and the END node presents the final workflow result. The workflow 1 consists of a single LLM agent with tool node. The workflow 2 is managed with a supervisor node, which decides the next acting agent, the research or the simulator. }
    \Description{}
    \label{fig:wfs}
\end{figure*}

We built two LangGraph workflows to handle different test runs, shown in \autoref{fig:wfs}. 
Here we used OpenAI/gpt-4o-mini as the LLM models.~\cite{gpt4omini_openai}
Workflow 1 (\autoref{fig:wfs}A) was a single simulation agent with tool node, that calls functions to download protein structure from Protein Data Bank, and run MD simulations. 
It's suitable for tasks with provided PDB ID in the user prompt. 
Workflow 2 (\autoref{fig:wfs}B) featured the supervisor agent to coordinate the flow of action, via deciding the next active agent, based on current graph state. 
In addition to the simulation agent, the workflow also incorporated a research agent with Tavily search tools~\cite{tavily_search}, that can query the relevant information from the internet to provide additional context for the LLM agents. 
This ability of querying the internet gave the workflow 2 to access the up-to-date knowledge. 
It was applied for more complex tasks, when extra information was required.

\subsection{Parsl: The HPC workload manager}
We adopted Parsl as the workload manager for the tool calls. 
Parsl is a Python library that allows the user to execute native Python functions in parallel and asynchronous workflows.~\cite{babuji19parsl, ward2021colmena}
It bridges user-defined Python functions and the computing resource via the Parsl configuration, requiring minimal code changes when moving across platforms, from a single laptop to leading-edge HPC systems. 
Parsl uses a dataflow programming model that is based on the function calls and input/output, thereby allowing the users to focus on the scientific workflow, instead of juggling the HPC computing queue. 
These features make Parsl the ideal intermediate for integrating the LLM tool calls and HPC resources. 

\subsection{Molecular dynamics (MD) simulation function}
We used molecular dynamics simulations to test the integration of the LangGraph tool calls and Parsl workload. 
The simulation function is set up with OpenMM software package to run on the Nvidia CUDA GPUs.~\cite{eastman2023openmm}
The function takes an input of a PDB file, and set up the simulation parameters, such as nonbonded cutoff distance, temperature, pressure for NPT system, timestep, and simulation length. 
In the LangGraph workflow, the LLM model calls this simulation function and assigns these parameters accordingly. 

In addition, the function cleans up the PDB file via pdbfixer, to add missing heavy atoms in the protein chain. 
It builds the system topology, which comprises all the atomic bonded and nonboned interaction, using Gromacs pdb2gmx~\cite{abraham2015gromacs} with Charm36m force field~\cite{huang2013charmm36}. 
The forcefield directory is defined in the environment variable as \emph{GMX\_ff}. 
The simulation timestep is default at 2 fs, and integrated with LangevinMiddleIntegrator. 
The long range nonbonded interactions are calculated with Particle Mesh Ewald method.~\cite{darden1993particle}

The MD simulation function also manages the simulation directories. 
Each time the function is called, it creates a new directory with the current time stamp and a unique ID to avoid the race condition of multiple parallel function creating the result directory at the same time. 
All the simulation results will be stored in their respective directories.

\subsection{LangChain tool call via Parsl}

\begin{figure}
    \centering
    \includegraphics[width=.85\linewidth]{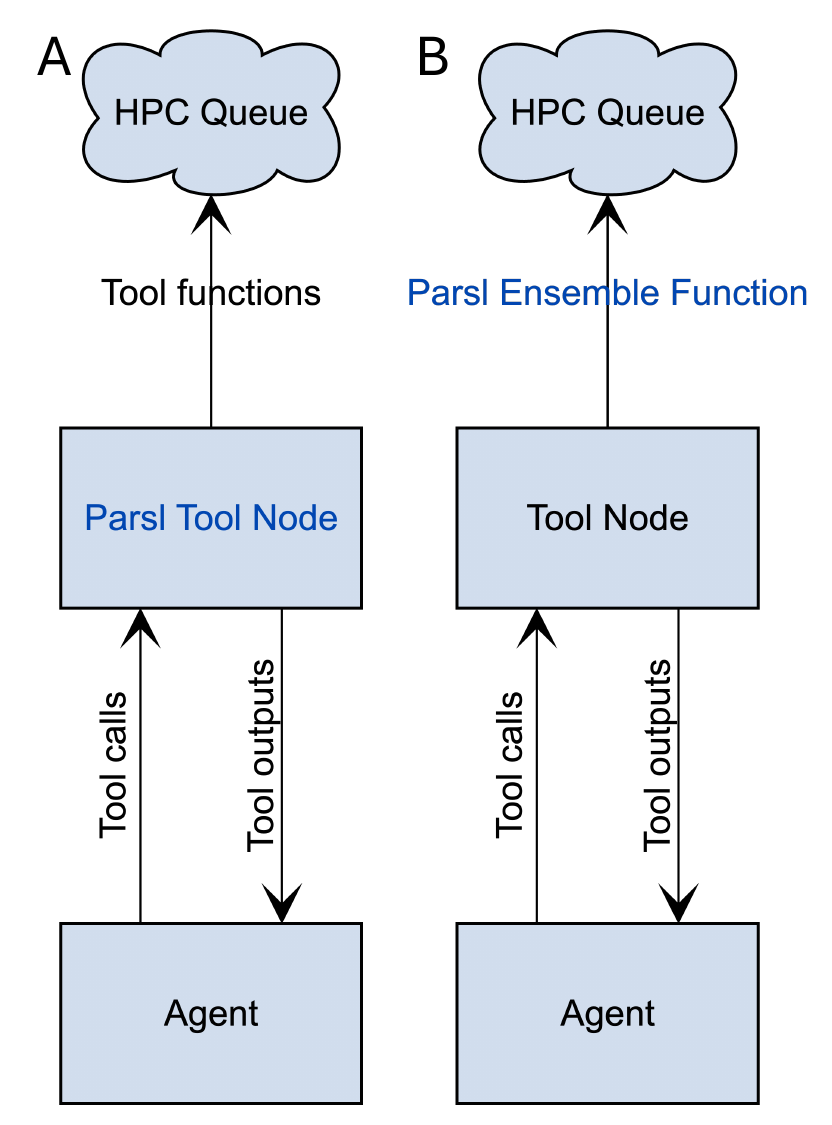}
    \caption{Two Parsl tool call scheme for LangGraph. The figure demonstrates the LLM agent, tool node and the tool functions. The Parsl tool node (A) replaces the LangGraph tool node, which submits the tool functions to the parallel Parsl queue. The Parsl ensemble function (B) can be directly called by the LLM, and launches an ensemble of simulation runs with Parsl. The blue letters mark the Parsl implementation to the LangGraph tool call framework. }
    \Description{}
    \label{fig:tool_call}
\end{figure}

We set up two different tool call integrations of Parsl to LangGraph shown in \autoref{fig:tool_call}, specifically,
\begin{enumerate}
    \item Parsl tool node: A tool node that launches functions via Parsl. 
    \item Parsl ensemble function: Tool function that runs an ensemble of tools with Parsl. 
\end{enumerate}

In setup  1 (\autoref{fig:tool_call}A), the Parsl tool node replaces the tool node of LangGraph in the workflow. 
When the LLM agent generates tool calls, it launches them to the Parsl queue, distributing them to the computing resource. 
The tool functions are executed concurrently by the Parsl workers. 
The tool node collects and returns the outputs to the LLM agent, once all the tasks are done. 

The setup 2 (\autoref{fig:tool_call}B) manages the multitasking through the tool function. 
The function embeds the Parsl to distribute multiple simulation runs to the computing system. 
During a tool call instance, the LLM model inputs the number of simulation runs to the function as an argument, and the function launches all the simulation runs to Parsl queue. 
This setup is more suitable to launch many simulation runs on the HPC queue, as the tool call cap of the LLM agent. 


\subsection{Experiment setup}
We prompted the LangGraph AI workflows of the previous section with inputs, shown in \autoref{tab:prompts}, to test the Parsl tool call functions. 
To test Parsl setup on different platforms, the workflows are set up for both local Nvidia GPU workstations and HPC system, Polaris/ALCF. 
The local workstation applied the tool call setup 1 with a Parsl-implemented tool node, which launched jobs to a Parsl queue. 
The HPC runs were more compatible with tool call setup 2, where the tool functions were implemented with Parsl to run simulation ensembles.

In the prompts, we specified the input protein structure with either a PDB ID directly, or the protein name. 
With prompts providing the PDB ID, workflow 1 was applied, as no additional information is needed to download the input protein.
Prompts with only protein names or acronym were inputted for the more complex workflow 2. 
Additionally, the simulation parameters were specified in the prompts, namely the number of runs, temperature and length. 
For all the local workstation runs, 8 simulation runs were specified in the prompts on 4 or 8 GPUs on the machine. 
On Polaris, we tested 100 simulations with 100 GPUs on 25 nodes and 80 runs with 40 GPUs on 10 nodes.

\begin{table*}
  \caption{Prompt setup for experiments}
  \label{tab:prompts}
  \begin{tabular}{p{0.05\linewidth}p{0.2\linewidth}p{0.1\linewidth}p{0.1\linewidth}p{0.45\linewidth}}
    \toprule
    Run & Platform & Tool Call & Workflow& Prompt\\
    \midrule
    \texttt{1} & workstation & 1 & 1 &Can you run 8 simulations on \{path of a local PDB file\} in 313 K for 50 ps? \\
    \texttt{2} & workstation & 1 & 1 &Can you find and download 2KKJ, and run 8 simulations on them in 313 K for 50 ps? \\
    \texttt{3}& workstation & 1 & 2 &Can you find the complex structure for NCBD/ACTR,  and run 8 simulations of them in 310 K for 100 ps? \\
    \texttt{4}& workstation& 1 & 2 &Can you find and download 8 crystal structures of lysozyme from PDB, and run simulations of them in 310 K for 100 ps? \\
    \texttt{5}& Polaris/100 GPUs& 2 & 1 &Can you download the structure of 2KKJ from Protein Data Bank, and run 100 simulations of it in 313 K for 50 ps?\\
    \bottomrule
  \end{tabular}
\end{table*}

%% file: results.tex
\section{Results}
The prompt results included both the LLM outputs and the simulation runs, spawned from the tool calls. 
Examining this information can gain us insights on the performance and limitations of these LLM agent workflows. 

\subsection{Local workstation}
The local runs were performed on a lambda machine with 8 Nvidia V100 GPUs. 
We prompted the workflow to run 8-simulation ensemble run of different protein input instructions, including a local PDB file, PDB ID, or just protein names. 
These runs implemented the Parsl tool node setup, where the tool node was in charge to submit the tasks to the Parsl queue.  

The run 1 and 2 were set up on the workflow 1 with a single LLM agent and simulation tools, and implemented with the Parsl tool node. 
The model was able to choose the correct simulation input parameters, including the number of tool calls, based on the input prompts. 
These 8 tool functions were then distributed to the 4 GPUs on the workstation via Parsl. 
As the result, both workflows generated 8 simulation runs of 50 ps in 313 K on its respective protein structure input, as the prompt specified. 

A local PDB file path was specified in the run 1 prompt as the input for the MD simulation. 
During setting up the prompt, we noticed the absolute file path on the file system was required for the tool function to locate the correct PDB file.  
The LLM agent recognized the full file path in the prompt and set up tool functions with correct arguments. 

The prompt 2 specified the PDB ID, 2KKJ, of nuclear coactivator binding domain (NCBD), instead of the file path. 
To obtain the PDB file, the LLM agent called the tool function to download the protein structure from PDB, based on the provided PDB ID. 
The PDB download function returned the log containing the PDB file path to the LLM agent, which then assigned the PDB file to the simulation tool functions.

\begin{figure*}
    \centering
    \includegraphics[width=.85\textwidth]{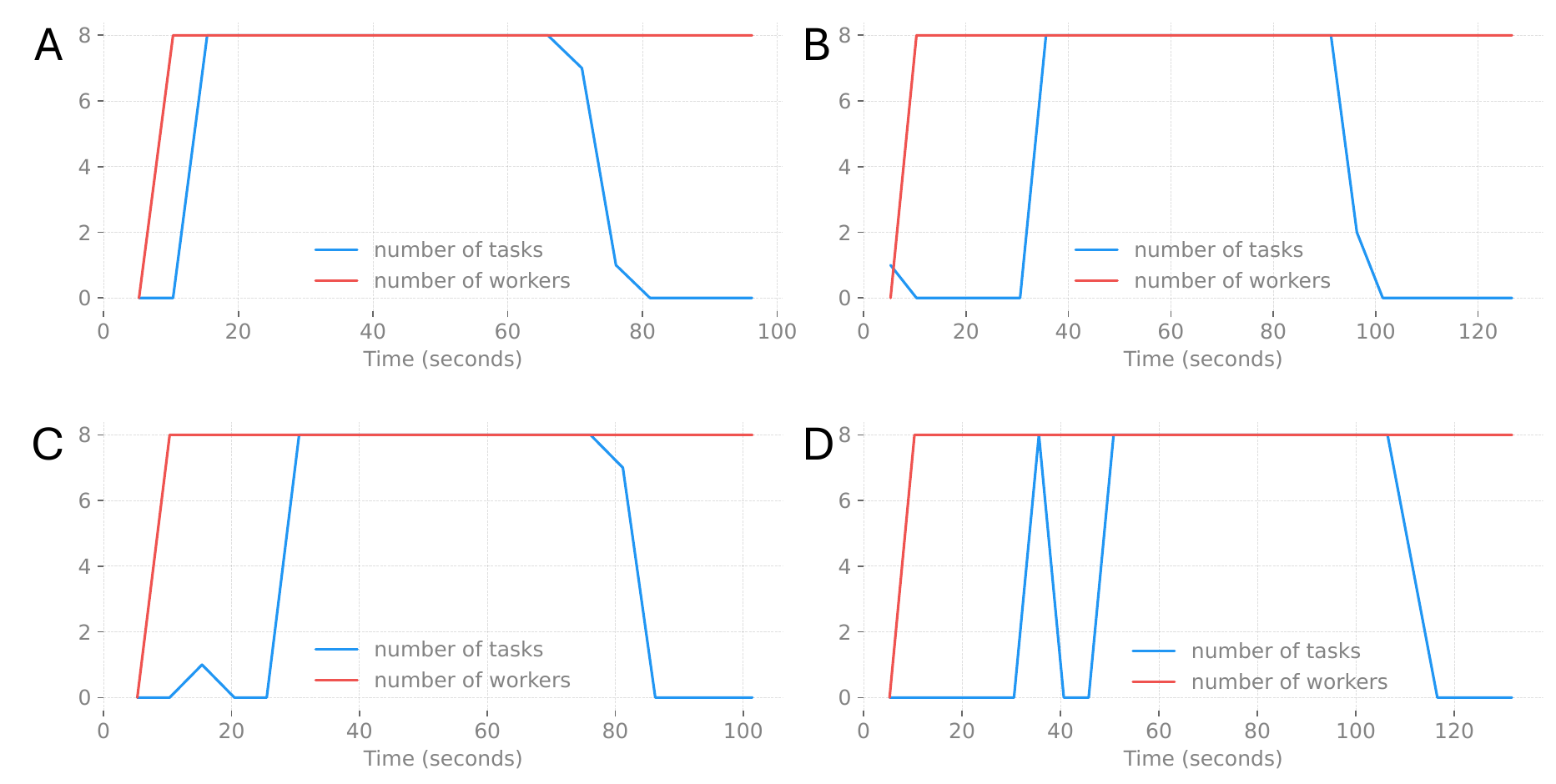}
    \caption{The Parsl process for the run 1 (A), 2 (B), 3 (C) and 4 (D). Both run launched 8 workers with GPUs. The Parsl logs were recorded every 5 seconds.  }
    \Description{}
    \label{fig:local_run}
\end{figure*}

The simulation runs of 2KKJ were run at $\sim$260 ns/day on Nvidia V100 GPUs. 
The Parsl run process were shown in \autoref{fig:local_run} A and B. 
The workers were initiated at $\sim$5 s, and the MD tasks were submitted at $\sim$10 seconds in run 1 with local PDB file. 
For run 2 with the PDB ID, the workflow first submitted the single task to download the PDB structure, and submitted the MD tasks at around $\sim$30 s. 
It took both runs $\sim$70 seconds to finish all 8 simulations, and 15–25 seconds for Parsl shutdown process. 

Compared to the run 1 and 2, prompt 3 and 4 contained the protein names, instead of local PDB file path or PDB ID. 
They required additional information to set up the simulation runs. 
Therefore, for these prompts, we applied the workflow 2, which comprises three LLM agents, supervisor, researcher and simulator. 
The input was directed by the supervisor to the researcher agent, which obtained the PDB ID based on the input prompt. 
The search tool took the refine query from researcher agent, and retrieved the 5 search results. 
These results were then refined with the LLM agent, and passed back to the supervisor. 
The next acting agent, simulator, submitted all the simulation runs to the Parsl worker queue. 

In the run 3, the LLM agent queried for a single protein complex structure of CBP nuclear coactivator binding domain and activator for thyroid hormone and retinoid receptors (NCBD/ACTR). 
The LLM agent made two tool calls via the Tavily search tool, with \emph{NCBD ACTR complex structure} and \emph{simulate NCBD ACTR 310 K 100 ps}, for the context before and after the \emph{and} in the prompt. 
From the \emph{NCBD ACTR complex structure} query, the search tool retrieved 5 results about the NCBD structure.~\cite{dogan2013transition_ncbd,marino2018charge_ncbd,chen2023structural_ncbd,bauer2021conformational_ncbd,karlsson2019structurally_ncbd}
These results covered studies of NCBD/ACTR complex in heterodimer interaction, binding/folding formation, and conformational editing, via X-ray diffraction (XRD), nuclear magnetic resonance (NMR), small angle neutron scattering, molecular simulations, et al. 
The 2nd query, \emph{NCBD ACTR 310 K 100 ps}, consisted of the protein complex name and the simulation temperature and length, and also returned 5 query results, including one of the query 1 result on molecular dynamics simulation~\cite{karlsson2019structurally_ncbd}. 
Noted, none of these results were the original paper for the NCBD/ACTR NMR structure, 1KBH~\cite{demarest2002mutual_1kbh}. 
But the PDB ID was referred in the context of the query results, and picked up by the research agent. 
It also summarized these results into a brief report with references about NCBD/ACTR complex structure, covering the NMR and XRD structures, binding characteristics, and simulation protocols for production. 

The supervisor agent then passed the task to the simulator. 
The simulator then downloaded the PDB structure from the RCSB database, and initiated 8  simulation runs with the Parsl worker. 
The run process was shown in \autoref{fig:local_run}C. 
The download tool was launched at $\sim$10 seconds for the 1KBH structure. 
The NCBD/ACTR simulations ran at $\sim$350 ns/day. 
All 8 simulation were then submitted to the Parsl workers and finished in $\sim$60 seconds. 
The results were printed as a list of the trajectories.

The run 4 queried 8 crystal structures of lysozyme. 
We noticed that the number of search results must be larger than 8, in order for the LLM agent to identify all 8 structures. 
When the search tool only returned the top 5 results, it fetched PDB IDs of 148L~\cite{kuroki1993covalent_148L}, 
1B7E~\cite{davies1999three_1b7e}, 
1LYZ, 
2LYZ~\cite{diamond1974real_1lyz}, 
3WEL~\cite{tagami2015structural_3wel}, 
5R2Z~\cite{wollenhaupt2020f2x_5r2z}, 
7BVM~\cite{nam2020polysaccharide_7bvm}, 
and 7F26~\cite{liang2021novel_7f26}.
Among these results, 2LYS is the partially-folded intermediate of CylR2, 3WEL is Sugar Beet $\alpha$-Glucosidase and 5R2Z is Endothiapepsin. 
These PDB IDs were hallucinated by the LLM agent, as only 5 search results were generated from the researcher. 
When retrieving 10 results with the search tool, the LLM agent downloaded the structures of 1LYZ, 
2LYZ, 
3LYZ, 
4LYZ, 
5LYZ, 
6LYZ~\cite{diamond1974real_1lyz}, 
7LYZ~\cite{herzberg1983protein_7lyz}, 
and 8LYZ~\cite{beddell1975x_8lyz}.
These PDB IDs were confirmed as lysozyme protein structures, providing correct inputs for the MD simulations. 
To avoid LLM hallucination, it required sufficient queried results from search tools than the requested information. 
Especially, in this case, the number of requested protein structures must be less than the number of returned results from researcher agent. 

Different from the previous runs, the workflow invoked 8 parallel download functions for all 8 PDB IDs. 
These functions were called at $\sim$30 seconds  into the workflow, due to the long query time for the more complex task of getting 8 PDB IDs. 
The simulations were then set up and run on the 8 Parsl workers for $\sim$70 seconds, at the speed of 310 ns/day. 
From the run with 5 queries, we received the wrong lysozyme entries, which had different protein structures from the correct lysozyme entries. 
We were able to observe the load imbalance between Parsl workers, where the smaller proteins were finished earlier than the larger ones, with $\sim$177 seconds difference in execution time. 

These runs on a local workstation showed the parallel execution of tool functions, especially those computationally expensive ones, such as MD simulation, can be implemented in the agent workflow to reduce execution time. 
The implementation of Parsl to the tool node packaged the LangChain tool functions and submitted the tool calls to the workers for parallel execution. 
This can reduce the overhead of the users designing the parallelism logic. Instead, they can focus on design of the tools and agent workflow.

\subsection{Polaris runs}
The HPC runs were carried out on Polaris/ALCF machine production queue. 
The tasks on Polaris system is managed with PBS system.  
We implemented Parsl to allocate the computing resource, and submit the LangGraph tool functions to the Polaris computing nodes. 
We applied the simple workflow 1, as the PDB ID was provided in the prompt. 

First, we tested the different tool call setups, and concluded the Parsl tool function, tool call setup 2, was more compatible for the HPC systems. 
While implemented with the Parsl tool node, the LLM agent made 24 tool calls, with \emph{100 simulations} specified in the input prompt during one experiment with Parsl tool node (setup 1). 
Therefore, we needed a different implementation on the HPC systems for massive parallel tool calls, to make sure the correct number of simulation runs were instantiated on the computing resource. 
Parsl was implemented with the simulation tool function to run an ensemble of simulation runs, instead of a Parsl tool node launching each individual simulation to the queue. 
The number of simulation runs was now an argument to the Parsl simulation ensemble function. 
The LLM agent only needed to pass the right number of simulations to this tool function. 
This implementation bypassed the LLM agent calling an excessive number of tools, making it more suitable for the HPC computing environment. 
Furthermore, the user can now decide which function would be running on the computing node, giving the flexibility to move only the computationally expensive tasks to the Parsl queue.  
However, it required the user to explicitly rewrite the tool function or add a wrapper function to incorporate the Parsl toolkit for the simulation ensemble runs. 

\begin{figure}
    \centering
    \includegraphics[width=.85\linewidth]{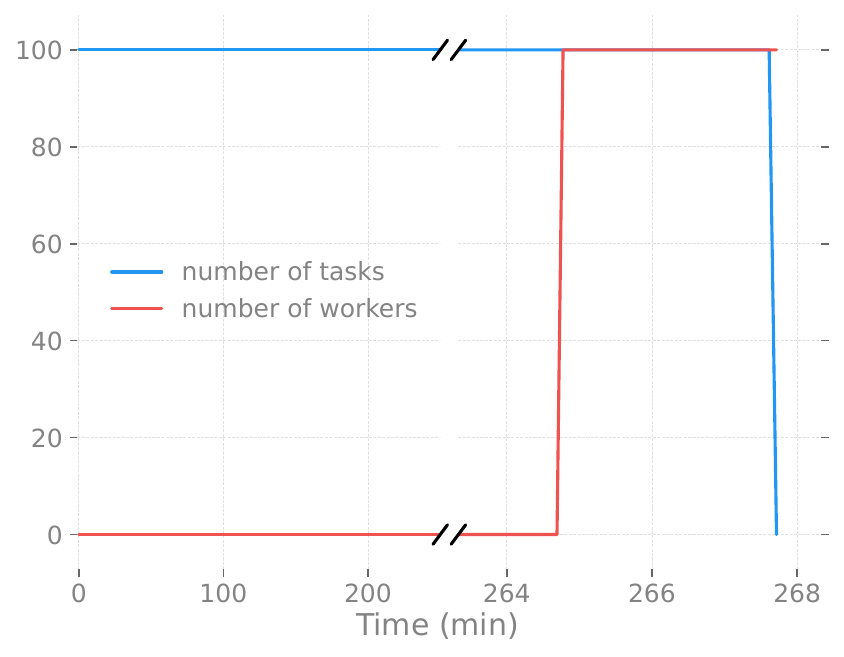}
    \caption{The timeline of Parsl simulation ensemble on Polaris, running 100 simulations of 2KKJ protein from LangGraph agent workflow. The number of workers represented the Parsl workers with available GPUs, and the number of tasks depicted the tasks in the Parsl queue. }
    \Description{}
    \label{fig:polaris_run}
\end{figure}

With the Parsl tool function, the LangGraph workflow was able to access the Polaris PBS queue directly without setting up the submission commands or scripts. 
The simulation tasks were submitted to the Parsl queue, which then assigned them to the workers with available resources.
The timeline of execution is shown in \autoref{fig:polaris_run}. 
The LLM agent identified and downloaded 2KKJ structure on the login node, and initiated the Parsl simulation function. 
This tool function launched 100 simulation to the Parsl queue. 
With pre-defined configuration, Parsl requested 25 nodes from the Polaris PBS. 
The tasks waited in the queue for $\sim$265 minutes, until 25 nodes were available on Polaris. 
The Parsl workers were assigned with GPUs and subsequently a task per worker. 
On the Nvidia A100 GPUs, the 50-ps simulations of 2KKJ ran in $\sim$340 ns/day, and were then finished  within $\sim$3 minutes. 
The majority of the run time was the queue time on Polaris. 
Depending on the HPC system, we could also separate the task into smaller chunks to reduce the queue time. 

From this result, the large set of tool execution was limited by the LLM agent tool call ability. 
We were able to circumvent this issue by combining the tool functions and Parsl to run the tool ensemble. 
These tool functions can be invoked with the LangGraph default tool node directly, however, required the user to implement the Parsl logic to the tool functions. 
At the meantime, the user has more control over where the functions will be executed, giving more versatility to the workflow.


%% file: conclusion.tex
\section{Conclusion}
This work aims to improve the tool execution of the LLM agent workflow for scientific applications, especially those requires heavy computing. 
Using MD simulation as the example task, we designed and tested two different Parsl implementation on both local workstation and HPC system. 
With Parsl, we can work on the high-level Python language, and express the parallelism with simple decorators and safe execution. 
The Parsl parallelism was implemented at different layers of tool execution in these two setups.  
It resulted in difference in the application, compatibility and ease-to-use between two implementation. 

The first implementation incorporated Parsl parallel execution in the tool node layer. 
Therefore, all the invoked tool functions were converted into a Parsl python function and submited to the workers. 
This setup requires no change to the tool functions. 
But all the tool functions that bound with the LLM agent, are executed with Parsl, including the download functions in our application. 
This is adequate for the workflow on the local workstation, where all tasks are carried out with local resource and a simple universal computing environment. 

Once implemented this setup on the Polaris system, we noticed the LLM agent had an upper call limits for the tools, and different system environments on the login and computing node also affected the workflow execution. 
For example, the computing node had no internet access by default, and the queue time was long on the HPC system despite the workload of the task. 
Therefore, it is unnecessary to execute all tools on the computing queue. 

We then developed the second implementation of Parsl ensemble tool function, where the Parsl parallelism was only set up for the expensive simulation functions. 
Instead of calling a function to run one simulation, the simulation function now invokes a simulation ensemble that ran multiple simulation concurrently. 
This gives more flexibilities of how and where the tasks are executed. 
However, it is now the user's responsibility to develop the simulation ensemble function with Parsl. 

We also implemented two workflows for scientific tasks with different requirements of external information. 
When exact information, such as file path or PDB ID for MD simulation, is present, the AI workflow can identify from the given prompts and invoke the simulation functions. 
Additional information is essential when only generic instruction is included in the prompt. 

Our implementation enables the LLM agent to access any compatible computing resource through Parsl.
Both setups add to the ability of LLM agent to call and assign right function arguments to the tools, of distributing the task execution via Parsl to the system computing resource (GPUs in our applications). 
Due to the LLM tool call limitation and computing system requirement, it still requires tailoring of the tool call setup depending on the tasks and computing platforms. 

Designing the tool function is another important requirement for the LLM agent workflow. 
The scientific application is the combination of flexible AI calls/executions and rigid step-by-step scientific protocol. 
The former enables the workflow to make decision on which function to call and assigning the right function arguments. 
The latter is well-defined in the practice of many scientific domains. 
For example, the MD simulation requires the setup of topology that includes all the molecular interactions. 
We can merge the topology building and simulation function, as they are always execution sequentially. 
This end-to-end design is less dependent on the LLM agents, and less likely to err, but too rigid for more sophisticated tasks. 
For the analysis step, it requires independent functions to carry out different methods, such RMSD, RMSF, radius of gyration, et al. 
The splitting of the scientific protocol into LLM agent tool steps is crucial for designing versatile LLM agent workflow for scientific applications. 

The future work will focus on how to make more capable LLM agent workflow for scientific research application. 
For example, AI-enhanced sampling of molecular simulations~\cite{ma2020deep_ddmd,brace2022coupling_ddmd,lee2019deepdrivemd_ddmd} can be implemented with Parsl-enabled LLM agent network, with more carefully designed workflows, agents and tools~\cite{nadeem2024optimizing_policy}. 
With Parsl to manage the computing tasks and resource, the workflow can leverage the current HPC platform, such as Aurora/ALCF, frontier/OLCF, etc. 
Furthermore, the human-in-loop design can also give the real-time user feedback or intervention to the workflow. 
We are also working on the expanding the capability of the workflow, by reducing the user interaction with Parsl, implementing function environment management, improving logging and checkpointing, and enabling asynchronous execution of different tool functions.
